\documentclass[aps,prl,onecolumn,superscriptaddress,showpacs,preprint]{revtex4-1}
\usepackage{amsmath,amsfonts,amssymb,graphicx}
\usepackage{dcolumn}
\usepackage{bm}
\usepackage{epstopdf}
\usepackage{subfigure}
\usepackage{multirow}
\usepackage[mathlines]{lineno}

\begin{document}
	
\title{A High Accuracy Electrical Stopping Power Prediction Model based on Deep Learning Algorithm and its Applications}

\author{\surname{Xun} Guo}
\affiliation{State Key Laboratory of Nuclear Physics and Technology, School of Physics, Peking University, Beijing 100871, P. R. China} 

\author{\surname{Hao} Wang}
\affiliation{State Key Laboratory of Nuclear Physics and Technology, School of Physics, Peking University, Beijing 100871, P. R. China} 
\affiliation{CAPT, HEDPS and IFSA Collaborative Innovation Center of MoE, College of Engineering, Peking University, Beijing 100871, P. R. China}

\author{\surname{Shijun} Zhao}
\affiliation{Department of Mechanical Engineering, City University of Hong Kong, Hong Kong, P. R. China}

\author{\surname{Ke} Jin}
\affiliation{State Key Laboratory of Nuclear Physics and Technology, School of Physics, Peking University, Beijing 100871, P. R. China} 
\email{jinke@bit.edu.cn}

\author{\surname{Jianming} Xue}
\affiliation{State Key Laboratory of Nuclear Physics and Technology, School of Physics, Peking University, Beijing 100871, P. R. China} 
\affiliation{CAPT, HEDPS and IFSA Collaborative Innovation Center of MoE, College of Engineering, Peking University, Beijing 100871, P. R. China}
\email{jmxue@pku.edu.cn}

\date{\today}

\begin{abstract}
Energy loss of energetic ions in solid is crucial in many field, and accurate prediction of the ion stopping power is a long-time goal. Though great efforts have been made, it is still very difficult to find a universal prediction model to accurately calculate the ion stopping power in distinct target materials. Deep learning algorithm is a newly emerged method to solve multi-factors physical problems and can mine the deeply implicit relations among parameters, which make it a powerful tool in energy loss prediction. In this work, we developed an energy loss prediction model based on deep learning. When experimental data are available, our model can give predictions with an average absolute difference close to 5.7\%, which is in the same level compared with other widely used programs e.g. SRIM. In the regime without experimental data, our model still can maintain a high performance, and has higher reliability compared with the existing models. The ion range of Au ions in SiC can be calculated with a relative error of 0.6$\sim$25\% for ions in the energy range of 700$\sim$10'000 keV, which is much better than the results calculated by SRIM. Moreover, our model support the reciprocity conjecture of ion stopping power in solid proposed by P. Sigmund, which has been known for a long time but can hardly been proved by any of the existing stopping power models. This high-accuracy energy loss prediction model is very important for the research of ion-solid interaction mechanism and enormous relevant applications of energetic ions, such as in semiconductor fabrications, nuclear energy systems and the space facilities. 

\end{abstract}

\maketitle

In the past century, the energy loss of energetic ions in matter has been significant subject which received great attention, since this topic is important no matter for the scientific researches or industrial applications, including radiation damage, material analysis by using ion beams, nuclear physics, ion implantation and nano-device modification \cite{arnau1990stopping, PhysRevLett.99.235501, PhysRevLett.108.225504, PhysRevLett.111.215002}. For these reasons, enormous experimental measurements and plenty of theoretical studies, by Bethe \cite{Bethe1930}, Bloch \cite{Bloch1933}, Lindhard \cite{Lindhard1954}, Sigmund \cite{Sigmund1982}, Ziegler \cite{ZIEGLER1988215} and et al., have been carried out to achieve a better description of ion stopping power, which can be defined as the energy loss in unit distance per atom. Meanwhile, many code, like SRIM \cite{SRIM1985, SRIM2008, SRIM2010} and MSTAR \cite{MSTAR2001, MSTAR2003}, have been developed to calculate these values and simulate the ion transportation behavior in solids.

Energy loss of energetic ions in solid is complicated: both the ion parameters and solid properties may affect the energy transfer process. Many researches have proved that the electronic stopping power is a complex function of collision parameters, including atomic number and mass for incident ion ($Z_1$, $M_1$) and target material ($Z_2$, $M_1$), along with the kinetic energy of incident ion ($E_{in}$). Besides, the contribution of other factors, such as binding energy, state of matter, electron structure, band gap and excitation energy, also have already been proved to be very important for the stopping process \cite{SALAH1998382, arnau1990stopping, Artacho2007, PhysRevLett.99.235501, PhysRevLett.79.4112}. 
However, nearly all the widely accepted empirical energy loss models only consider the influence of collision parameters during the fitting and optimizing of these models, and therefore, their prediction accuracy and transferability are fundamentally restricted. Moreover, limited by the ion source technique and measuring difficulties, the experimental data of some specific ion types or energy ranges are nearly impossible to be measured by far, which means the prediction accuracy can hardly been further improved by increasing the experimental database for these existing methods based on semi-theoretical formulas. A new method which can overcome these problems has been desired for many years.

Deep Learning (DL) is a new computational method which is composed of multiple processing layers to represent data with multi-levels of abstraction \cite{lecun2015deep}. It is widely believed to be good at solving multi-parameters problems and mining the deeply implicit relations among these data. The algorithm of DL has already brought about breakthroughs in many fields of science, e.g. planning chemical syntheses \cite{Segler2018}, acceleration of super-resolution localization microscopy \cite{Ouyang2018, Nehme2018, Strack2018}, classifying scientific data \cite{webb2018deep, PhysRevLett.120.141103} and solving high-dimensional problems in condensed matter system \cite{Carrasquilla2017, Carleo2017, vanNieuwenburg2017, Xia2018, Torlai2018}. Considering the large number of energy loss data obtained in the past decades, it can be expected that DL based model can provide higher accuracy and better transferability than traditional methods, such as SRIM and MSTAR, which are strongly relay on some artificially selected stopping power formulas or highly approximate fitting functions.

In this letter, we demonstrated that DL provides a good breakthrough point to inroad nuclear technology and particle physics. The proposed Deep Learning based Electronic Stopping Power (DL-ESP) model can provide good accuracy, whose average relative absolute difference between predict and experimental results can reach 5.7\%. More importantly, this model shows much higher accuracy at regime without experimental data than the existing models. Furthermore, DL-ESP model provides evidence of the so call reciprocity principle proposed by P. Sigmund \cite{Sigmund2008}, and also can give a much better prediction in evaluating the projected range of Au ion in SiC. 

To deploy DL-ESP, we used the experimental ion collision data collected by H. Paul and International Atomic Energy Agency (IAEA), which contain nearly all the published electronic stopping power data of different ion in solids, gases and compounds, since the 1990's \cite{Paul-IAEA, MONTANARI201750}. 16'907 experimental data of energy loss of energetic ions, whose kinetic energy between 5 keV/u to 100 MeV, in elementary substance were selected, and 13'526 (80\%) of them were randomly chosen as training database, while the rest 3'381 (20\%) were used for validation. Besides, in order to verify the DP-ESP model still can make prediction accurately while the experimental data for a certain ion are totally absent, all the experimental data of Ti ion and Kr ion in Al, were excluded from the training database. To simplify the problem, we also do not consider the crystal structure of target materials in the current work.

\begin{figure}
 \centering
 \includegraphics[width=0.9\textwidth]{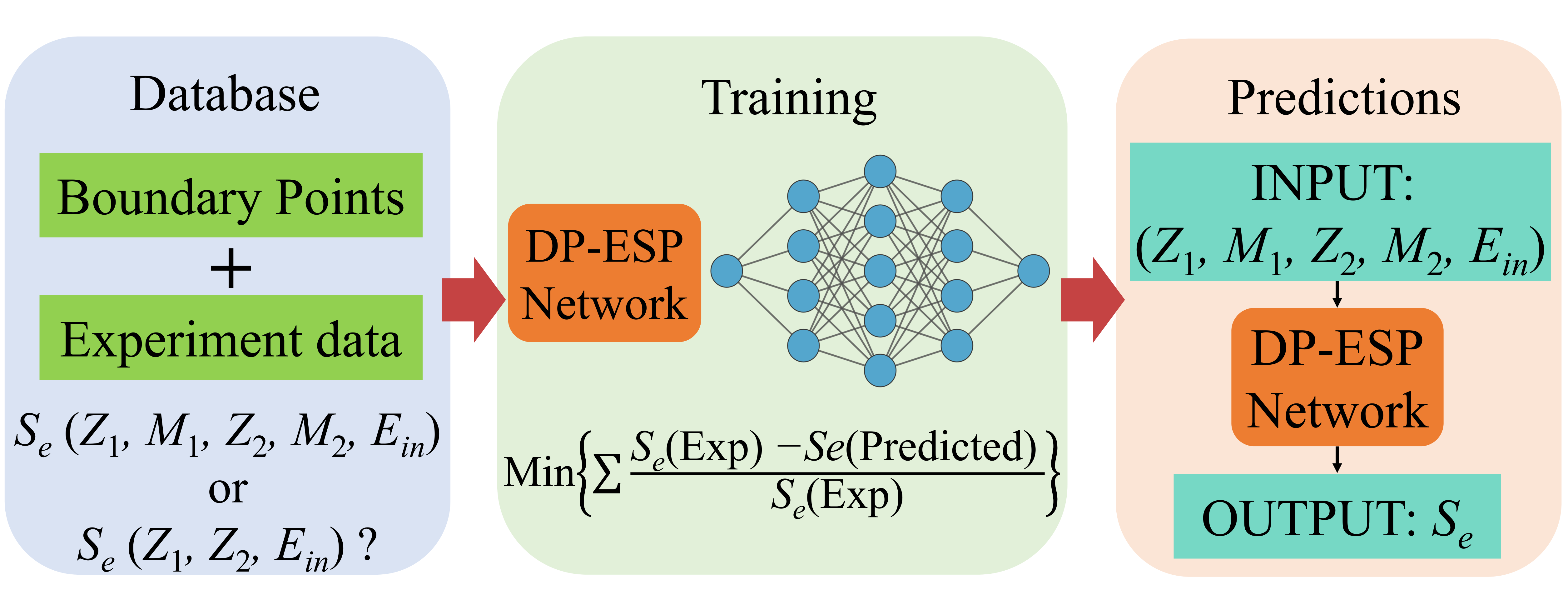}
 \caption{Schematic of DL-ESP from the atomic collision database and how to use it performing a highly accurate prediction. This procedure involved training data set of evaluated experimental data and boundary points.}
 \label{fig1}
\end{figure}

Following previous studies, we assumed that the electronic stopping power ($S_e$) is only related with  $Z_1$, $M_1$, $Z_2$, $M_2$ and $E_{in}$. According to our test, a 5 layer network with grid of 16$\times$32$\times$64$\times$32$\times$16 is good enough to give the required accuracy. All the DL-ESP model, including the generating, training and invoking of neural network, was built based on Keras \cite{chollet2015keras} with TensorFlow \cite{tensorflow2015} platform as the back-end. In order to balance the prediction performance for $E_{in}$ at both high and low energy range, the mean absolute percentage error (MAPE) loss function was adopted, instead of the mean absolute error (MAE) loss functions. Moreover, we chose the rectified linear unit (ReLU) activate function \cite{agarap2018deep} and the adaptive moment estimation (Adam) optimizer \cite{kingma2014adam} to make the network easier to be converged. We export the DL-ESP model which has the smallest validate loss during the training process, and invoked it at the following exploration and discussion. A simple illustration of our method can be shown as FIG. \ref{fig1}. 

Since most of the experimental ion (except H and He) stopping power lack of low energy data, 12'800 boundary points were also included in training database, in order to maintain the accuracy of boundary at extreme low energy range. These points were generated by using LSS model, which is widely believed to be accurate enough for low-velocity ion through matter \cite{Lindhard1954, Lindhard1961, Lindhard1963, Sigmund1983LSS}:
\begin{equation}
\label{eq1}
S_e=3.83\dfrac{Z_1^{7/6}Z_2}{M_1^{1/2}\left(Z_1^{2/3}+Z_2^{2/3}\right)^{3/2}}E_{in}^{1/2}
\end{equation}
in which the unit of $S_e$ and $E_{in}$ is $10^{-15}$ eV$\text{cm}^2$/atom and keV. By using Eq. (\ref{eq1}), $S_e$ of 80 different kinds of ions, with the incident energy of 1 and 10 keV, in 80 types of elementary materials were calculated, then used as boundary points only in training database. In the following study, the results of SRIM were used as a benchmark to compare with the accuracy of DP-ESP, because SRIM is the most widely used simulation software for the stopping and range of ions in materials. Moreover, we also investigated the influence of input parameters, by exploring a simplified version of DL-ESP (DL-ESP-Simple) with the minimum input parameters ($Z_1$, $Z_2$, $E_{in}$). A detailed comparison between these two models was discussed in Supplementary Material.

\begin{table*}[h]
	\centering
	\caption{A brief comparison of DL-ESP and SRIM, by exhibiting the average absolute relative difference ($\bar{\sigma}$) between predicted $S_e$ and experimental results of H, He, Li and other heavy ions, and the general consistency within 5\% and 10\%, which represent the percentage of absolute relative difference ($\left| \sigma \right|$) which is smaller than 5\% and 10\%, for the experimental data used in this work.}	
	\begin{ruledtabular}
		\begin{tabular}{lccccc}
			                                 & H ions & He ions & Li ions & Other ions & Overall \\ 
		\hline
			     $N_{Data}$ (DL-ESP)      & 8'748  &  6'127  &   291   &   1'741    & 16'907  \\	
			     $N_{Data}$ (SRIM)       & 8'300  &  6'500  &  1'400  &   9'000    & 25'200  \\	        
			         $\bar{\sigma}$ (DL-ESP-Simple)         & 7.1\%  &  7.1\%  &  4.5\%  &   10.7\%    &  7.5\%  \\
			       $\bar{\sigma}$ (DL-ESP)         & 5.6\%  &  6.0\%  &  4.2\%  &   7.0\%    &  5.7\%  \\
			       $\bar{\sigma}$ (SRIM)          & 4.0\%  &  3.9\%  &  4.8\%  &   5.8\%    &  4.6\%  \\		 
			 $\left| \sigma \right|$ $\textless$ 5\% (DL-ESP-Simple)  &  63\%  &  61\%   &  70\%   &    46\%    &  53\%   \\
			$\left| \sigma \right|$ $\textless$ 5\% (DL-ESP)  &  69\%  &  69\%   &  73\%   &    64\%    &  69\%   \\
			$\left| \sigma \right|$ $\textless$ 5\% (SRIM)   &  74\%  &  76\%   &  72\%   &    58\%    &  69\%   \\
			$\left| \sigma \right|$ $\textless$ 10\% (DL-ESP-Simple) &  81\%  &  82\%   &  88\%   &    67\%    &  73\% \\
			$\left| \sigma \right|$ $\textless$ 10\% (DL-ESP) &  85\%  &  86\%   &  89\%   &    83\%    &  86\%	\\
			$\left| \sigma \right|$ $\textless$ 10\% (SRIM)  &  87\%  &  89\%   &  83\%   &    82\%    &  86\%   \\
		\end{tabular}
	\end{ruledtabular}
\label{table1}
\end{table*}

As shown in Table I, the average absolute relative difference ($\bar{\sigma}$) of DL-ESP is about 5.7\%, quite close to the performance of SRIM (4.6\%), and better than MSTAR (about 1$\sim$2\% at high energy, but for low energy ions it increase to 10$\sim$20\% \cite{MSTAR2003}). As for the general consistency within 5\% and 10\%, which can be defined as the percentage of experimental data whose predicted value with absolute relative difference ($\left| \sigma \right|$) smaller than 5\% and 10\%, SRIM and DL-ESP have similar performance, especially for incident ion heavier than He. 



\begin{figure}
	\centering
	\includegraphics[width=0.9\textwidth]{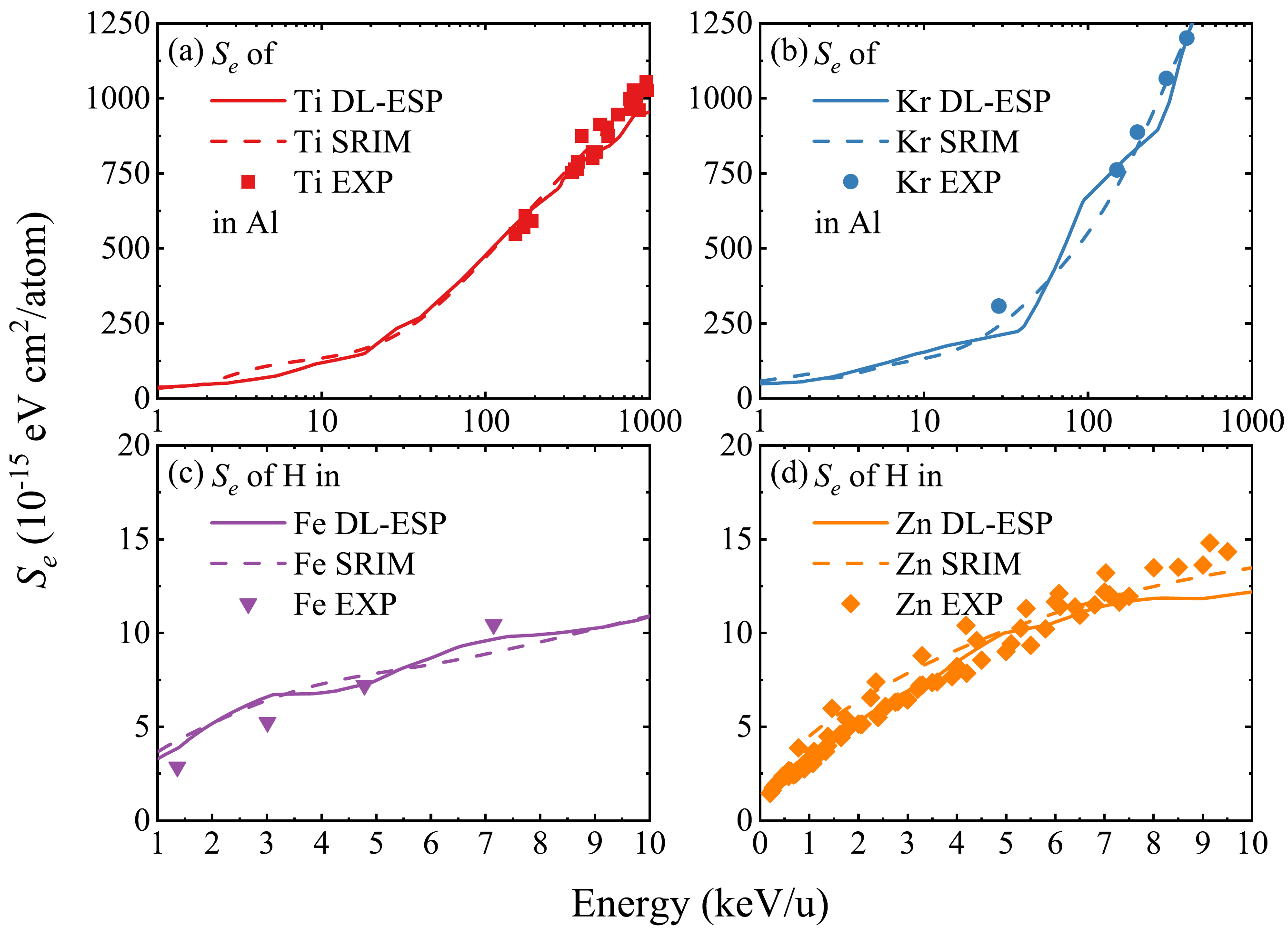}
	\caption{A brief illustration of the expansion capability of DL-ESP. The electronic stopping power of (a) Ti and (b) Kr in Al, as well as H in (c) Fe and (d) Zn at low energy range were calculated, and verified by using the extra experimental data and SRIM.}
	\label{fig2}
\end{figure}

To verify the transferability of DL-ESP, we calculated the predicted $S_e$ for some experimental points which were excluded from the entire database, especially the data of medium-heavy ions, as shown in FIG. \ref{fig2}. It could be demonstrated from FIG. \ref{fig2} that DL-ESP exhibit extraordinary agreement for these extra experimental results, such as Ti and Kr in Al, along with additional data of H in Fe and Zn as representatives. Although these data points were not included in neither training nor validate database, this model can provide the prediction curve with high accuracy only based on nearby data, which means its transferability can be greatly improved if more training data with abundant atomic and energy information were introduced into the system. 



As discussed before, this model is established based on LSS model (boundary points for low energy ions) and experimental data (for ions' $E_{in} > $ 5 keV/u), so it is also very important to evaluate the necessity and accuracy of this treatment. FIG. \ref{fig2} (c) and (d) can illustrate the ability of extension of this model at low energy range, especially when 1 keV/u $< E_{in} <$ 10 keV/u. Though we didn't import any experimental data (whose $E_{in}$ is smaller than 5 keV/u) into the deep learning network, DL-ESP can still give an accurate enough prediction at this energy range, with high consistency compared with both SRIM and experimental results. Therefore it can be concluded from FIG. \ref{fig2} that the DL-ESP model has the ability of extrapolation to a certain extent, with the help of boundary point generated by LSS model, and this treatment is necessary and helpful when there is no enough experimental data for low energy ions.

\begin{figure}
	\centering
	\includegraphics[width=0.9\textwidth]{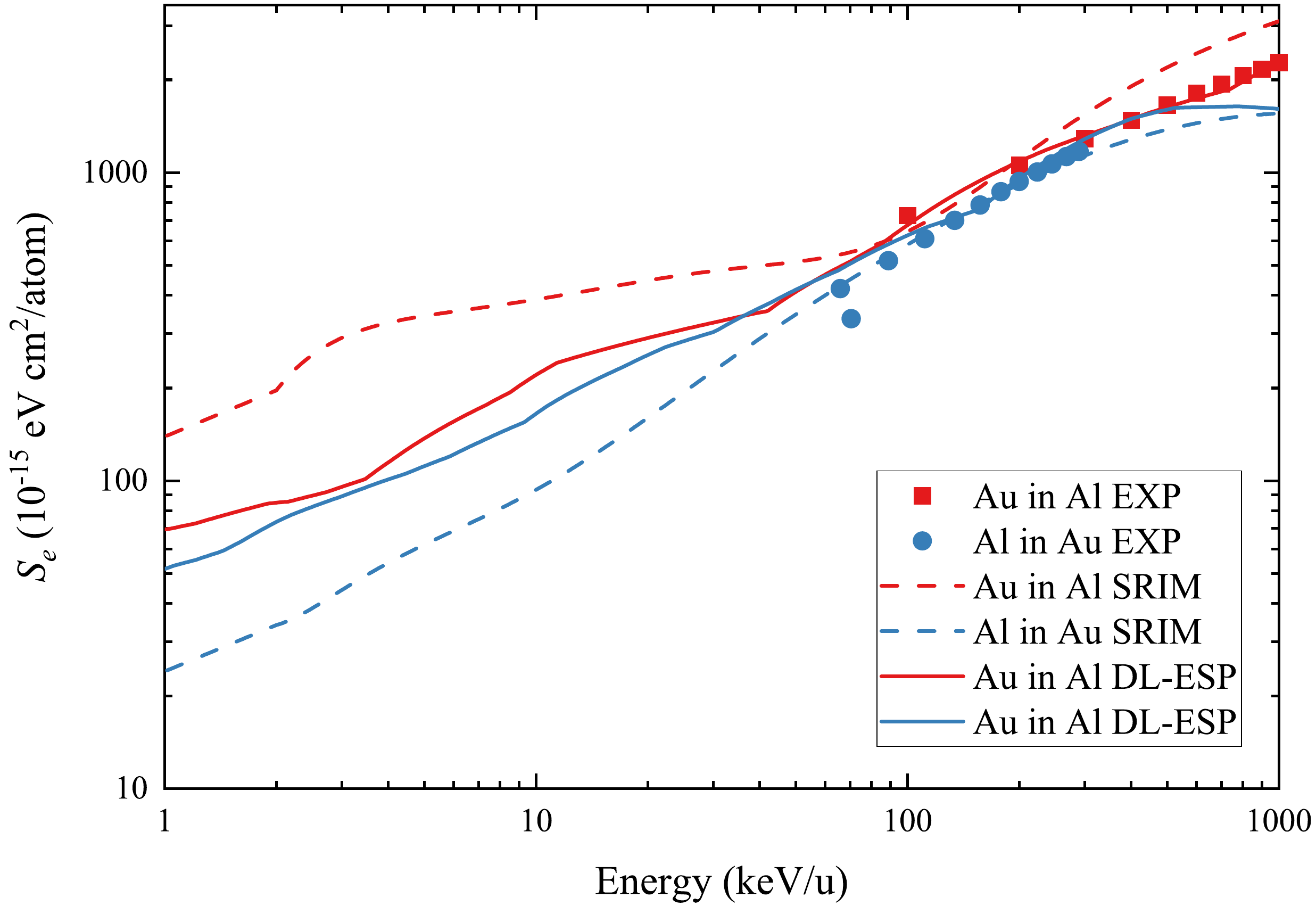}
	\caption{According to SRIM's calculation, there is an obvious difference between $S_e$(Au in Al) and $S_e$(Al in Au) at low energy range. While for DL-ESP, the predicted $S_e$ curves are basically coincident, and the $S_e$ values at low energy range are significantly lower than SRIM's results. }
	\label{fig3}
\end{figure}

Moreover, it can be shown from TABLE \ref{table1} that the data of heavy ions only took a relatively small percentage of total database, so the accuracy of heavy ions need to be carefully verified. In fact, other calculation models, such as SRIM and MSTAR, also face the similar problem: it has been already discussed from other researchers that the electronic stopping cross section of some ultra-heavy ion, e.g. Au, is overestimated by SRIM when $E_{in}$ near 1000 keV/u \cite{SIGMUND1998}, as illustrated in FIG. \ref{fig3}. The abnormal high $S_e$ at extreme low energy (about 140 in the unit of $10^{-15}$ eV$\text{cm}^2$/atom for 1 keV/u Au ion in Al) is also quite controversial, because the low energy models of heavy ions used in SRIM are simply extrapolate the fitting curve of light ions, which would cause obvious deviation systematically. As for DL-ESP, the network would establish extrapolation based on the nearby data points, so the behavior of $S_e$ curve for Au ion at low energy range is much more reasonable than SRIM.


It should also be noted that the high accuracy of SRIM is based on segment fitting functions for different mass and energy range, so it can provide well consistence with existing experimental results. However, for the regime that lack the support of data, the accuracy of these fitting functions becomes quite controversial, because they may neglect the influence of other potential factors. According to the discussion in Supplementary Material, the tiny difference between DL-ESP and DL-ESP-Simple reveals that DL network can indeed mining the inline relations deeply hiding in these experimental data, which could significantly enhance the transferability of this model. Therefore, though DL-ESP can hardly achieve higher overall accuracy than SRIM, its $\bar{\sigma}$ is a macroscopic evaluation of the entire database equally for each atom types and energy ranges. In other words, DL-ESP would possess excellent transferability since it would have similar accuracy for the conditions which are shortage of experimental data theoretically. 


FIG. \ref{fig3} also plotted the $S_e$ curve of Al in Au, to demonstrate Sigmund's reciprocity theory, in which he believed that when ion's velocity smaller than Bohr's velocity ($c/$137 = 2.088$\times$10$^{6}$ m/s, about 22.75 keV/u), the electronic stopping power of $A$ ion in $B$ material is equal to $B$ ion in $A$: $S_e(A~\text{in}~B)$ = $S_e(B~\text{in}~A)$ \cite{Sigmund2008}. So far, this conjecture is actually correct for all the existing experimental results, but it can hardly be proved theoretically or experimentally, due to the limit of experimental techniques. However, it is widely accepted to use Sigmund's reciprocity theory to evaluate the rationality of newly emerged results. This DL-ESP model can support Sigmund's conjecture, and a more detailed investigation about reciprocity theory is explored in the Supplementary Material.

\begin{figure}
	\centering
	\includegraphics[width=0.9\textwidth]{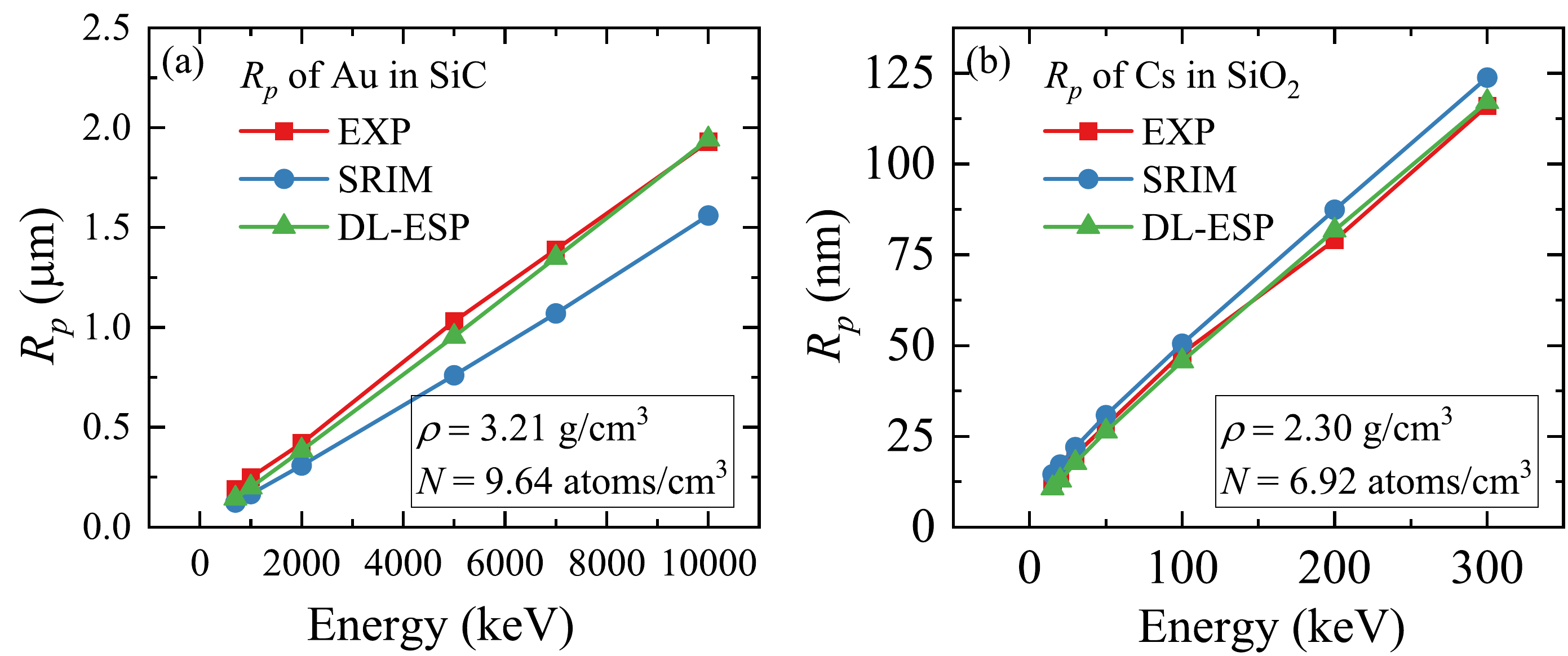}
	\caption{The projected range ($R_p$) of Au in SiC predicted by SRIM and DL-ESP. It should be note that DL-ESP cannot obtain $R_p$ directly. This results are calculated with the help of Lindhard and Bragg's theory.}
	\label{fig4}
\end{figure}

To further demonstrate the accuracy for DL-ESP, we systemically calculated the projected range ($R_p$) by DL-ESP and SRIM, compared with experimental results, as shown in FIG. \ref{fig4}. To minimize the influence of nuclear stopping power ($S_n$), the $S_n$ used in this work is also calculated by SRIM, so the difference exhibited in FIG. \ref{fig4} only represents the contribution of $S_e$. The calculation of ion range by using DL-ESP model is mainly based on the Bragg's theory \cite{Bragg1905} (to convert $S_e$ of elementary materials into compounds) and Lindhard formula \cite{Lindhard1968} (to evaluate $R_p$ by using $S_e$ and $S_n$), detail information about the calculation has also been illustrated in Supplementary Material.


It can be shown in FIG. \ref{fig4} that, the predicted $R_p$ from DL-ESP and SRIM are all very close to the experimental result when $E_{in}$ is small enough. However, when $E_{in}$ becomes greater, the DL-ESP's results could still agree well with the experimental data, while $R_p$ from SRIM calculation became less accurate compared with DL-ESP: the $R_p$ of 700$\sim$5000 keV Au in SiC predicted by SRIM are about 20$\sim$40\% lower than experimental results \cite{JIN201365}, and for high energy range, the difference becomes more evident, mainly due to the incorrectly estimated stopping power. Similar deviation can also be seen in the result of Cs in SiO$_2$ \cite{GRANDE198817}.


In conclusion, we developed a new electronic stopping power model based on deep learning method to describe ion's electronic energy loss in elementary materials. Compared with traditional method, deep learning can extensively avoid the systematic error generated by fitting model, capture more detailed information of experimental data, and completely eradicate the interference of subjective factors. All these features makes DL-ESP exhibit high accuracy and good transferability even with small size of database. This model can not only increase the prediction accuracy of ion projected range, but also provide a new insight to challenge the classical puzzles, such as the reciprocity conjecture which can hardly be verified experimentally or theoretically. These findings should be of great importance for fundamentally understanding the novel physical phenomena related to the ion interaction with materials, and could be extensively necessary for vast relevant areas, such as semiconductor industry and nano-structures preparation.
 
This work is financially supported by NSFC (Grant No. 11705010) and the China Postdoctoral Science Foundation (Grant No. 2019M650351). We are grateful for computing resource provided by Weiming No. 1 and Life Science No. 1 High Performance Computing Platform at Peking University, and TianHe-1(A) at National Supercomputer Center in Tianjin.



\begin{thebibliography}{44}%
	\makeatletter
	\providecommand \@ifxundefined [1]{%
		\@ifx{#1\undefined}
	}%
	\providecommand \@ifnum [1]{%
		\ifnum #1\expandafter \@firstoftwo
		\else \expandafter \@secondoftwo
		\fi
	}%
	\providecommand \@ifx [1]{%
		\ifx #1\expandafter \@firstoftwo
		\else \expandafter \@secondoftwo
		\fi
	}%
	\providecommand \natexlab [1]{#1}%
	\providecommand \enquote  [1]{``#1''}%
	\providecommand \bibnamefont  [1]{#1}%
	\providecommand \bibfnamefont [1]{#1}%
	\providecommand \citenamefont [1]{#1}%
	\providecommand \href@noop [0]{\@secondoftwo}%
	\providecommand \href [0]{\begingroup \@sanitize@url \@href}%
	\providecommand \@href[1]{\@@startlink{#1}\@@href}%
	\providecommand \@@href[1]{\endgroup#1\@@endlink}%
	\providecommand \@sanitize@url [0]{\catcode `\\12\catcode `\$12\catcode
		`\&12\catcode `\#12\catcode `\^12\catcode `\_12\catcode `\%12\relax}%
	\providecommand \@@startlink[1]{}%
	\providecommand \@@endlink[0]{}%
	\providecommand \url  [0]{\begingroup\@sanitize@url \@url }%
	\providecommand \@url [1]{\endgroup\@href {#1}{\urlprefix }}%
	\providecommand \urlprefix  [0]{URL }%
	\providecommand \Eprint [0]{\href }%
	\providecommand \doibase [0]{http://dx.doi.org/}%
	\providecommand \selectlanguage [0]{\@gobble}%
	\providecommand \bibinfo  [0]{\@secondoftwo}%
	\providecommand \bibfield  [0]{\@secondoftwo}%
	\providecommand \translation [1]{[#1]}%
	\providecommand \BibitemOpen [0]{}%
	\providecommand \bibitemStop [0]{}%
	\providecommand \bibitemNoStop [0]{.\EOS\space}%
	\providecommand \EOS [0]{\spacefactor3000\relax}%
	\providecommand \BibitemShut  [1]{\csname bibitem#1\endcsname}%
	\let\auto@bib@innerbib\@empty
	\bibitem [{\citenamefont {Arnau}\ \emph {et~al.}(1990)\citenamefont {Arnau},
		\citenamefont {Pealba}, \citenamefont {Echenique}, \citenamefont {Flores},\
		and\ \citenamefont {Ritchie}}]{arnau1990stopping}%
	\BibitemOpen
	\bibfield  {author} {\bibinfo {author} {\bibfnamefont {A.}~\bibnamefont
			{Arnau}}, \bibinfo {author} {\bibfnamefont {M.}~\bibnamefont {Pealba}},
		\bibinfo {author} {\bibfnamefont {P.~M.}\ \bibnamefont {Echenique}}, \bibinfo
		{author} {\bibfnamefont {F.}~\bibnamefont {Flores}}, \ and\ \bibinfo {author}
		{\bibfnamefont {R.~H.}\ \bibnamefont {Ritchie}},\ }\href@noop {} {\bibfield
		{journal} {\bibinfo  {journal} {Physical Review Letters}\ }\textbf {\bibinfo
			{volume} {65}},\ \bibinfo {pages} {1024} (\bibinfo {year}
		{1990})}\BibitemShut {NoStop}%
	\bibitem [{\citenamefont {Pruneda}\ \emph {et~al.}(2007)\citenamefont
		{Pruneda}, \citenamefont {S\'anchez-Portal}, \citenamefont {Arnau},
		\citenamefont {Juaristi},\ and\ \citenamefont
		{Artacho}}]{PhysRevLett.99.235501}%
	\BibitemOpen
	\bibfield  {author} {\bibinfo {author} {\bibfnamefont {J.~M.}\ \bibnamefont
			{Pruneda}}, \bibinfo {author} {\bibfnamefont {D.}~\bibnamefont
			{S\'anchez-Portal}}, \bibinfo {author} {\bibfnamefont {A.}~\bibnamefont
			{Arnau}}, \bibinfo {author} {\bibfnamefont {J.~I.}\ \bibnamefont {Juaristi}},
		\ and\ \bibinfo {author} {\bibfnamefont {E.}~\bibnamefont {Artacho}},\ }\href
	{\doibase 10.1103/PhysRevLett.99.235501} {\bibfield  {journal} {\bibinfo
			{journal} {Phys. Rev. Lett.}\ }\textbf {\bibinfo {volume} {99}},\ \bibinfo
		{pages} {235501} (\bibinfo {year} {2007})}\BibitemShut {NoStop}%
	\bibitem [{\citenamefont {Zeb}\ \emph {et~al.}(2012)\citenamefont {Zeb},
		\citenamefont {Kohanoff}, \citenamefont {S\'anchez-Portal}, \citenamefont
		{Arnau}, \citenamefont {Juaristi},\ and\ \citenamefont
		{Artacho}}]{PhysRevLett.108.225504}%
	\BibitemOpen
	\bibfield  {author} {\bibinfo {author} {\bibfnamefont {M.~A.}\ \bibnamefont
			{Zeb}}, \bibinfo {author} {\bibfnamefont {J.}~\bibnamefont {Kohanoff}},
		\bibinfo {author} {\bibfnamefont {D.}~\bibnamefont {S\'anchez-Portal}},
		\bibinfo {author} {\bibfnamefont {A.}~\bibnamefont {Arnau}}, \bibinfo
		{author} {\bibfnamefont {J.~I.}\ \bibnamefont {Juaristi}}, \ and\ \bibinfo
		{author} {\bibfnamefont {E.}~\bibnamefont {Artacho}},\ }\href {\doibase
		10.1103/PhysRevLett.108.225504} {\bibfield  {journal} {\bibinfo  {journal}
			{Phys. Rev. Lett.}\ }\textbf {\bibinfo {volume} {108}},\ \bibinfo {pages}
		{225504} (\bibinfo {year} {2012})}\BibitemShut {NoStop}%
	\bibitem [{\citenamefont {Grabowski}\ \emph {et~al.}(2013)\citenamefont
		{Grabowski}, \citenamefont {Surh}, \citenamefont {Richards}, \citenamefont
		{Graziani},\ and\ \citenamefont {Murillo}}]{PhysRevLett.111.215002}%
	\BibitemOpen
	\bibfield  {author} {\bibinfo {author} {\bibfnamefont {P.~E.}\ \bibnamefont
			{Grabowski}}, \bibinfo {author} {\bibfnamefont {M.~P.}\ \bibnamefont {Surh}},
		\bibinfo {author} {\bibfnamefont {D.~F.}\ \bibnamefont {Richards}}, \bibinfo
		{author} {\bibfnamefont {F.~R.}\ \bibnamefont {Graziani}}, \ and\ \bibinfo
		{author} {\bibfnamefont {M.~S.}\ \bibnamefont {Murillo}},\ }\href {\doibase
		10.1103/PhysRevLett.111.215002} {\bibfield  {journal} {\bibinfo  {journal}
			{Phys. Rev. Lett.}\ }\textbf {\bibinfo {volume} {111}},\ \bibinfo {pages}
		{215002} (\bibinfo {year} {2013})}\BibitemShut {NoStop}%
	\bibitem [{\citenamefont {Bethe}(1930)}]{Bethe1930}%
	\BibitemOpen
	\bibfield  {author} {\bibinfo {author} {\bibfnamefont {H.}~\bibnamefont
			{Bethe}},\ }\href {\doibase 10.1002/andp.19303970303} {\bibfield  {journal}
		{\bibinfo  {journal} {Annalen der Physik}\ }\textbf {\bibinfo {volume}
			{397}},\ \bibinfo {pages} {325} (\bibinfo {year} {1930})}\BibitemShut
	{NoStop}%
	\bibitem [{\citenamefont {Bloch}(1933)}]{Bloch1933}%
	\BibitemOpen
	\bibfield  {author} {\bibinfo {author} {\bibfnamefont {F.}~\bibnamefont
			{Bloch}},\ }\href {\doibase 10.1002/andp.19334080303} {\bibfield  {journal}
		{\bibinfo  {journal} {Annalen der Physik}\ }\textbf {\bibinfo {volume}
			{408}},\ \bibinfo {pages} {285} (\bibinfo {year} {1933})}\BibitemShut
	{NoStop}%
	\bibitem [{\citenamefont {Lindhard}(1954)}]{Lindhard1954}%
	\BibitemOpen
	\bibfield  {author} {\bibinfo {author} {\bibfnamefont {J.}~\bibnamefont
			{Lindhard}},\ }\href@noop {} {\bibfield  {journal} {\bibinfo  {journal} {Dan.
				Vid. Selsk Mat.-Fys. Medd.}\ }\textbf {\bibinfo {volume} {28, no. 8}},\
		\bibinfo {pages} {41 } (\bibinfo {year} {1954})}\BibitemShut {NoStop}%
	\bibitem [{\citenamefont {Sigmund}(1982)}]{Sigmund1982}%
	\BibitemOpen
	\bibfield  {author} {\bibinfo {author} {\bibfnamefont {P.}~\bibnamefont
			{Sigmund}},\ }\href {\doibase 10.1103/PhysRevA.26.2497} {\bibfield  {journal}
		{\bibinfo  {journal} {Phys. Rev. A}\ }\textbf {\bibinfo {volume} {26}},\
		\bibinfo {pages} {2497} (\bibinfo {year} {1982})}\BibitemShut {NoStop}%
	\bibitem [{\citenamefont {Ziegler}\ and\ \citenamefont
		{Manoyan}(1988)}]{ZIEGLER1988215}%
	\BibitemOpen
	\bibfield  {author} {\bibinfo {author} {\bibfnamefont {J.}~\bibnamefont
			{Ziegler}}\ and\ \bibinfo {author} {\bibfnamefont {J.}~\bibnamefont
			{Manoyan}},\ }\href {\doibase https://doi.org/10.1016/0168-583X(88)90273-X}
	{\bibfield  {journal} {\bibinfo  {journal} {Nuclear Instruments and Methods
				in Physics Research Section B: Beam Interactions with Materials and Atoms}\
		}\textbf {\bibinfo {volume} {35}},\ \bibinfo {pages} {215 } (\bibinfo {year}
		{1988})}\BibitemShut {NoStop}%
	\bibitem [{\citenamefont {Ziegler~J.F.}(1985)}]{SRIM1985}%
	\BibitemOpen
	\bibfield  {author} {\bibinfo {author} {\bibfnamefont {B.~J.}\ \bibnamefont
			{Ziegler~J.F.}},\ }\href {\doibase
		https://doi.org/10.1007/978-1-4615-8103-1_3} {\emph {\bibinfo {title} {The
				Stopping and Range of Ions in Matter. In: Bromley D.A. (eds) Treatise on
				Heavy-Ion Scienc}}}\ (\bibinfo  {publisher} {Springer, Boston, MA},\ \bibinfo
	{year} {1985})\BibitemShut {NoStop}%
	\bibitem [{\citenamefont {Ziegler}\ \emph {et~al.}(2008)\citenamefont
		{Ziegler}, \citenamefont {Biersack},\ and\ \citenamefont
		{Ziegler}}]{SRIM2008}%
	\BibitemOpen
	\bibfield  {author} {\bibinfo {author} {\bibfnamefont {J.~F.}\ \bibnamefont
			{Ziegler}}, \bibinfo {author} {\bibfnamefont {J.~P.}\ \bibnamefont
			{Biersack}}, \ and\ \bibinfo {author} {\bibfnamefont {M.~D.}\ \bibnamefont
			{Ziegler}},\ }\href {http://lib.ugent.be/catalog/rug01:001467757} {\emph
		{\bibinfo {title} {SRIM : the stopping and range of ions in matter}}}\
	(\bibinfo  {publisher} {Chester (Md.) : SRIM},\ \bibinfo {year}
	{2008})\BibitemShut {NoStop}%
	\bibitem [{\citenamefont {Ziegler}\ \emph {et~al.}(2010)\citenamefont
		{Ziegler}, \citenamefont {Ziegler},\ and\ \citenamefont
		{Biersack}}]{SRIM2010}%
	\BibitemOpen
	\bibfield  {author} {\bibinfo {author} {\bibfnamefont {J.~F.}\ \bibnamefont
			{Ziegler}}, \bibinfo {author} {\bibfnamefont {M.}~\bibnamefont {Ziegler}}, \
		and\ \bibinfo {author} {\bibfnamefont {J.}~\bibnamefont {Biersack}},\ }\href
	{\doibase https://doi.org/10.1016/j.nimb.2010.02.091} {\bibfield  {journal}
		{\bibinfo  {journal} {Nuclear Instruments and Methods in Physics Research
				Section B: Beam Interactions with Materials and Atoms}\ }\textbf {\bibinfo
			{volume} {268}},\ \bibinfo {pages} {1818 } (\bibinfo {year} {2010})},\
	\bibinfo {note} {19th International Conference on Ion Beam
		Analysis}\BibitemShut {NoStop}%
	\bibitem [{\citenamefont {Paul}\ and\ \citenamefont
		{Schinner}(2001)}]{MSTAR2001}%
	\BibitemOpen
	\bibfield  {author} {\bibinfo {author} {\bibfnamefont {H.}~\bibnamefont
			{Paul}}\ and\ \bibinfo {author} {\bibfnamefont {A.}~\bibnamefont
			{Schinner}},\ }\href {\doibase https://doi.org/10.1016/S0168-583X(01)00576-6}
	{\bibfield  {journal} {\bibinfo  {journal} {Nuclear Instruments and Methods
				in Physics Research Section B: Beam Interactions with Materials and Atoms}\
		}\textbf {\bibinfo {volume} {179}},\ \bibinfo {pages} {299 } (\bibinfo {year}
		{2001})}\BibitemShut {NoStop}%
	\bibitem [{\citenamefont {Paul}\ and\ \citenamefont
		{Schinner}(2003)}]{MSTAR2003}%
	\BibitemOpen
	\bibfield  {author} {\bibinfo {author} {\bibfnamefont {H.}~\bibnamefont
			{Paul}}\ and\ \bibinfo {author} {\bibfnamefont {A.}~\bibnamefont
			{Schinner}},\ }\href {\doibase https://doi.org/10.1016/j.adt.2003.08.003}
	{\bibfield  {journal} {\bibinfo  {journal} {Atomic Data and Nuclear Data
				Tables}\ }\textbf {\bibinfo {volume} {85}},\ \bibinfo {pages} {377 }
		(\bibinfo {year} {2003})}\BibitemShut {NoStop}%
	\bibitem [{\citenamefont {Salah}\ \emph {et~al.}(1998)\citenamefont {Salah},
		\citenamefont {Touchrift},\ and\ \citenamefont {Saad}}]{SALAH1998382}%
	\BibitemOpen
	\bibfield  {author} {\bibinfo {author} {\bibfnamefont {H.}~\bibnamefont
			{Salah}}, \bibinfo {author} {\bibfnamefont {B.}~\bibnamefont {Touchrift}}, \
		and\ \bibinfo {author} {\bibfnamefont {M.}~\bibnamefont {Saad}},\ }\href
	{\doibase https://doi.org/10.1016/S0168-583X(98)00031-7} {\bibfield
		{journal} {\bibinfo  {journal} {Nuclear Instruments and Methods in Physics
				Research Section B: Beam Interactions with Materials and Atoms}\ }\textbf
		{\bibinfo {volume} {139}},\ \bibinfo {pages} {382 } (\bibinfo {year}
		{1998})}\BibitemShut {NoStop}%
	\bibitem [{\citenamefont {Artacho}(2007)}]{Artacho2007}%
	\BibitemOpen
	\bibfield  {author} {\bibinfo {author} {\bibfnamefont {E.}~\bibnamefont
			{Artacho}},\ }\href {\doibase 10.1088/0953-8984/19/27/275211} {\bibfield
		{journal} {\bibinfo  {journal} {Journal of Physics: Condensed Matter}\
		}\textbf {\bibinfo {volume} {19}},\ \bibinfo {pages} {275211} (\bibinfo
		{year} {2007})}\BibitemShut {NoStop}%
	\bibitem [{\citenamefont {Eder}\ \emph {et~al.}(1997)\citenamefont {Eder},
		\citenamefont {Semrad}, \citenamefont {Bauer}, \citenamefont {Golser},
		\citenamefont {Maier-Komor}, \citenamefont {Aumayr}, \citenamefont
		{Pe\~nalba}, \citenamefont {Arnau}, \citenamefont {Ugalde},\ and\
		\citenamefont {Echenique}}]{PhysRevLett.79.4112}%
	\BibitemOpen
	\bibfield  {author} {\bibinfo {author} {\bibfnamefont {K.}~\bibnamefont
			{Eder}}, \bibinfo {author} {\bibfnamefont {D.}~\bibnamefont {Semrad}},
		\bibinfo {author} {\bibfnamefont {P.}~\bibnamefont {Bauer}}, \bibinfo
		{author} {\bibfnamefont {R.}~\bibnamefont {Golser}}, \bibinfo {author}
		{\bibfnamefont {P.}~\bibnamefont {Maier-Komor}}, \bibinfo {author}
		{\bibfnamefont {F.}~\bibnamefont {Aumayr}}, \bibinfo {author} {\bibfnamefont
			{M.}~\bibnamefont {Pe\~nalba}}, \bibinfo {author} {\bibfnamefont
			{A.}~\bibnamefont {Arnau}}, \bibinfo {author} {\bibfnamefont {J.~M.}\
			\bibnamefont {Ugalde}}, \ and\ \bibinfo {author} {\bibfnamefont {P.~M.}\
			\bibnamefont {Echenique}},\ }\href {\doibase 10.1103/PhysRevLett.79.4112}
	{\bibfield  {journal} {\bibinfo  {journal} {Phys. Rev. Lett.}\ }\textbf
		{\bibinfo {volume} {79}},\ \bibinfo {pages} {4112} (\bibinfo {year}
		{1997})}\BibitemShut {NoStop}%
	\bibitem [{\citenamefont {Lecun}\ \emph {et~al.}(2015)\citenamefont {Lecun},
		\citenamefont {Bengio},\ and\ \citenamefont {Hinton}}]{lecun2015deep}%
	\BibitemOpen
	\bibfield  {author} {\bibinfo {author} {\bibfnamefont {Y.}~\bibnamefont
			{Lecun}}, \bibinfo {author} {\bibfnamefont {Y.}~\bibnamefont {Bengio}}, \
		and\ \bibinfo {author} {\bibfnamefont {G.~E.}\ \bibnamefont {Hinton}},\
	}\href@noop {} {\bibfield  {journal} {\bibinfo  {journal} {Nature}\ }\textbf
		{\bibinfo {volume} {521}},\ \bibinfo {pages} {436} (\bibinfo {year}
		{2015})}\BibitemShut {NoStop}%
	\bibitem [{\citenamefont {Segler}\ \emph {et~al.}(2018)\citenamefont {Segler},
		\citenamefont {Preuss},\ and\ \citenamefont {Waller}}]{Segler2018}%
	\BibitemOpen
	\bibfield  {author} {\bibinfo {author} {\bibfnamefont {M.~H.~S.}\
			\bibnamefont {Segler}}, \bibinfo {author} {\bibfnamefont {M.}~\bibnamefont
			{Preuss}}, \ and\ \bibinfo {author} {\bibfnamefont {M.~P.}\ \bibnamefont
			{Waller}},\ }\href {\doibase 10.1038/nature25978} {\bibfield  {journal}
		{\bibinfo  {journal} {Nature}\ }\textbf {\bibinfo {volume} {555}},\ \bibinfo
		{pages} {604} (\bibinfo {year} {2018})}\BibitemShut {NoStop}%
	\bibitem [{\citenamefont {Ouyang}\ \emph {et~al.}(2018)\citenamefont {Ouyang},
		\citenamefont {Aristov}, \citenamefont {Lelek}, \citenamefont {Hao},\ and\
		\citenamefont {Zimmer}}]{Ouyang2018}%
	\BibitemOpen
	\bibfield  {author} {\bibinfo {author} {\bibfnamefont {W.}~\bibnamefont
			{Ouyang}}, \bibinfo {author} {\bibfnamefont {A.}~\bibnamefont {Aristov}},
		\bibinfo {author} {\bibfnamefont {M.}~\bibnamefont {Lelek}}, \bibinfo
		{author} {\bibfnamefont {X.}~\bibnamefont {Hao}}, \ and\ \bibinfo {author}
		{\bibfnamefont {C.}~\bibnamefont {Zimmer}},\ }\href {\doibase
		10.1038/nbt.4106} {\bibfield  {journal} {\bibinfo  {journal} {Nature
				Biotechnology}\ }\textbf {\bibinfo {volume} {36}},\ \bibinfo {pages} {460}
		(\bibinfo {year} {2018})}\BibitemShut {NoStop}%
	\bibitem [{\citenamefont {Nehme}\ \emph {et~al.}(2018)\citenamefont {Nehme},
		\citenamefont {Weiss}, \citenamefont {Michaeli},\ and\ \citenamefont
		{Shechtman}}]{Nehme2018}%
	\BibitemOpen
	\bibfield  {author} {\bibinfo {author} {\bibfnamefont {E.}~\bibnamefont
			{Nehme}}, \bibinfo {author} {\bibfnamefont {L.~E.}\ \bibnamefont {Weiss}},
		\bibinfo {author} {\bibfnamefont {T.}~\bibnamefont {Michaeli}}, \ and\
		\bibinfo {author} {\bibfnamefont {Y.}~\bibnamefont {Shechtman}},\ }\href
	{\doibase 10.1364/OPTICA.5.000458} {\bibfield  {journal} {\bibinfo  {journal}
			{Optica}\ }\textbf {\bibinfo {volume} {5}},\ \bibinfo {pages} {458} (\bibinfo
		{year} {2018})}\BibitemShut {NoStop}%
	\bibitem [{\citenamefont {Strack}(2018)}]{Strack2018}%
	\BibitemOpen
	\bibfield  {author} {\bibinfo {author} {\bibfnamefont {R.}~\bibnamefont
			{Strack}},\ }\href {\doibase 10.1038/s41592-018-0028-9} {\bibfield  {journal}
		{\bibinfo  {journal} {Nature Methods}\ }\textbf {\bibinfo {volume} {15}},\
		\bibinfo {pages} {403} (\bibinfo {year} {2018})}\BibitemShut {NoStop}%
	\bibitem [{\citenamefont {Webb}(2018)}]{webb2018deep}%
	\BibitemOpen
	\bibfield  {author} {\bibinfo {author} {\bibfnamefont {S.}~\bibnamefont
			{Webb}},\ }\href@noop {} {\bibfield  {journal} {\bibinfo  {journal} {Nature}\
		}\textbf {\bibinfo {volume} {554}},\ \bibinfo {pages} {555} (\bibinfo {year}
		{2018})}\BibitemShut {NoStop}%
	\bibitem [{\citenamefont {Gabbard}\ \emph {et~al.}(2018)\citenamefont
		{Gabbard}, \citenamefont {Williams}, \citenamefont {Hayes},\ and\
		\citenamefont {Messenger}}]{PhysRevLett.120.141103}%
	\BibitemOpen
	\bibfield  {author} {\bibinfo {author} {\bibfnamefont {H.}~\bibnamefont
			{Gabbard}}, \bibinfo {author} {\bibfnamefont {M.}~\bibnamefont {Williams}},
		\bibinfo {author} {\bibfnamefont {F.}~\bibnamefont {Hayes}}, \ and\ \bibinfo
		{author} {\bibfnamefont {C.}~\bibnamefont {Messenger}},\ }\href {\doibase
		10.1103/PhysRevLett.120.141103} {\bibfield  {journal} {\bibinfo  {journal}
			{Phys. Rev. Lett.}\ }\textbf {\bibinfo {volume} {120}},\ \bibinfo {pages}
		{141103} (\bibinfo {year} {2018})}\BibitemShut {NoStop}%
	\bibitem [{\citenamefont {Carrasquilla}\ and\ \citenamefont
		{Melko}(2017)}]{Carrasquilla2017}%
	\BibitemOpen
	\bibfield  {author} {\bibinfo {author} {\bibfnamefont {J.}~\bibnamefont
			{Carrasquilla}}\ and\ \bibinfo {author} {\bibfnamefont {R.~G.}\ \bibnamefont
			{Melko}},\ }\href {\doibase 10.1038/nphys4035} {\bibfield  {journal}
		{\bibinfo  {journal} {Nature Physics}\ }\textbf {\bibinfo {volume} {13}},\
		\bibinfo {pages} {431} (\bibinfo {year} {2017})}\BibitemShut {NoStop}%
	\bibitem [{\citenamefont {Carleo}\ and\ \citenamefont
		{Troyer}(2017)}]{Carleo2017}%
	\BibitemOpen
	\bibfield  {author} {\bibinfo {author} {\bibfnamefont {G.}~\bibnamefont
			{Carleo}}\ and\ \bibinfo {author} {\bibfnamefont {M.}~\bibnamefont
			{Troyer}},\ }\href {\doibase 10.1126/science.aag2302} {\bibfield  {journal}
		{\bibinfo  {journal} {Science}\ }\textbf {\bibinfo {volume} {355}},\ \bibinfo
		{pages} {602} (\bibinfo {year} {2017})}\BibitemShut {NoStop}%
	\bibitem [{\citenamefont {van Nieuwenburg}\ \emph {et~al.}(2017)\citenamefont
		{van Nieuwenburg}, \citenamefont {Liu},\ and\ \citenamefont
		{Huber}}]{vanNieuwenburg2017}%
	\BibitemOpen
	\bibfield  {author} {\bibinfo {author} {\bibfnamefont {E.~P.~L.}\
			\bibnamefont {van Nieuwenburg}}, \bibinfo {author} {\bibfnamefont {Y.-H.}\
			\bibnamefont {Liu}}, \ and\ \bibinfo {author} {\bibfnamefont {S.~D.}\
			\bibnamefont {Huber}},\ }\href {\doibase 10.1038/nphys4037} {\bibfield
		{journal} {\bibinfo  {journal} {Nature Physics}\ }\textbf {\bibinfo {volume}
			{13}},\ \bibinfo {pages} {435} (\bibinfo {year} {2017})}\BibitemShut
	{NoStop}%
	\bibitem [{\citenamefont {Xia}\ and\ \citenamefont {Kais}(2018)}]{Xia2018}%
	\BibitemOpen
	\bibfield  {author} {\bibinfo {author} {\bibfnamefont {R.}~\bibnamefont
			{Xia}}\ and\ \bibinfo {author} {\bibfnamefont {S.}~\bibnamefont {Kais}},\
	}\href {\doibase 10.1038/s41467-018-06598-z} {\bibfield  {journal} {\bibinfo
			{journal} {Nature Communications}\ }\textbf {\bibinfo {volume} {9}},\
		\bibinfo {pages} {4195} (\bibinfo {year} {2018})}\BibitemShut {NoStop}%
	\bibitem [{\citenamefont {Torlai}\ \emph {et~al.}(2018)\citenamefont {Torlai},
		\citenamefont {Mazzola}, \citenamefont {Carrasquilla}, \citenamefont
		{Troyer}, \citenamefont {Melko},\ and\ \citenamefont {Carleo}}]{Torlai2018}%
	\BibitemOpen
	\bibfield  {author} {\bibinfo {author} {\bibfnamefont {G.}~\bibnamefont
			{Torlai}}, \bibinfo {author} {\bibfnamefont {G.}~\bibnamefont {Mazzola}},
		\bibinfo {author} {\bibfnamefont {J.}~\bibnamefont {Carrasquilla}}, \bibinfo
		{author} {\bibfnamefont {M.}~\bibnamefont {Troyer}}, \bibinfo {author}
		{\bibfnamefont {R.}~\bibnamefont {Melko}}, \ and\ \bibinfo {author}
		{\bibfnamefont {G.}~\bibnamefont {Carleo}},\ }\href {\doibase
		10.1038/s41567-018-0048-5} {\bibfield  {journal} {\bibinfo  {journal} {Nature
				Physics}\ }\textbf {\bibinfo {volume} {14}},\ \bibinfo {pages} {447}
		(\bibinfo {year} {2018})}\BibitemShut {NoStop}%
	\bibitem [{\citenamefont {Sigmund}(2008)}]{Sigmund2008}%
	\BibitemOpen
	\bibfield  {author} {\bibinfo {author} {\bibfnamefont {P.}~\bibnamefont
			{Sigmund}},\ }\href {\doibase 10.1140/epjd/e2008-00011-9} {\bibfield
		{journal} {\bibinfo  {journal} {The European Physical Journal D}\ }\textbf
		{\bibinfo {volume} {47}},\ \bibinfo {pages} {45} (\bibinfo {year}
		{2008})}\BibitemShut {NoStop}%
	\bibitem [{\citenamefont {Paul}(1990)}]{Paul-IAEA}%
	\BibitemOpen
	\bibfield  {author} {\bibinfo {author} {\bibfnamefont {H.}~\bibnamefont
			{Paul}},\ }\href@noop {} {\enquote {\bibinfo {title} {Stopping power for
				light ions},}\ } (\bibinfo {year} {1990}),\ \bibinfo {note} {available
		online. In: https://www-nds.iaea.org/stopping/}\BibitemShut {NoStop}%
	\bibitem [{\citenamefont {Montanari}\ and\ \citenamefont
		{Dimitriou}(2017)}]{MONTANARI201750}%
	\BibitemOpen
	\bibfield  {author} {\bibinfo {author} {\bibfnamefont {C.}~\bibnamefont
			{Montanari}}\ and\ \bibinfo {author} {\bibfnamefont {P.}~\bibnamefont
			{Dimitriou}},\ }\href {\doibase https://doi.org/10.1016/j.nimb.2017.03.138}
	{\bibfield  {journal} {\bibinfo  {journal} {Nuclear Instruments and Methods
				in Physics Research Section B: Beam Interactions with Materials and Atoms}\
		}\textbf {\bibinfo {volume} {408}},\ \bibinfo {pages} {50 } (\bibinfo {year}
		{2017})},\ \bibinfo {note} {proceedings of the 18th International Conference
		on the Physics of Highly Charged Ions (HCI-2016), Kielce, Poland, 11-16
		September 2016}\BibitemShut {NoStop}%
	\bibitem [{\citenamefont {Chollet}\ \emph {et~al.}(2015)\citenamefont {Chollet}
		\emph {et~al.}}]{chollet2015keras}%
	\BibitemOpen
	\bibfield  {author} {\bibinfo {author} {\bibfnamefont {F.}~\bibnamefont
			{Chollet}} \emph {et~al.},\ }\href@noop {} {\enquote {\bibinfo {title}
			{Keras},}\ }\bibinfo {howpublished} {\url{https://keras.io}} (\bibinfo {year}
	{2015})\BibitemShut {NoStop}%
	\bibitem [{\citenamefont {Abadi}\ \emph {et~al.}(2015)\citenamefont {Abadi},
		\citenamefont {Agarwal}, \citenamefont {Barham}, \citenamefont {Brevdo},
		\citenamefont {Chen}, \citenamefont {Citro}, \citenamefont {Corrado},
		\citenamefont {Davis}, \citenamefont {Dean}, \citenamefont {Devin},
		\citenamefont {Ghemawat}, \citenamefont {Goodfellow}, \citenamefont {Harp},
		\citenamefont {Irving}, \citenamefont {Isard}, \citenamefont {Jia},
		\citenamefont {Jozefowicz}, \citenamefont {Kaiser}, \citenamefont {Kudlur},
		\citenamefont {Levenberg}, \citenamefont {Man\'{e}}, \citenamefont {Monga},
		\citenamefont {Moore}, \citenamefont {Murray}, \citenamefont {Olah},
		\citenamefont {Schuster}, \citenamefont {Shlens}, \citenamefont {Steiner},
		\citenamefont {Sutskever}, \citenamefont {Talwar}, \citenamefont {Tucker},
		\citenamefont {Vanhoucke}, \citenamefont {Vasudevan}, \citenamefont
		{Vi\'{e}gas}, \citenamefont {Vinyals}, \citenamefont {Warden}, \citenamefont
		{Wattenberg}, \citenamefont {Wicke}, \citenamefont {Yu},\ and\ \citenamefont
		{Zheng}}]{tensorflow2015}%
	\BibitemOpen
	\bibfield  {author} {\bibinfo {author} {\bibfnamefont {M.}~\bibnamefont
			{Abadi}}, \bibinfo {author} {\bibfnamefont {A.}~\bibnamefont {Agarwal}},
		\bibinfo {author} {\bibfnamefont {P.}~\bibnamefont {Barham}}, \bibinfo
		{author} {\bibfnamefont {E.}~\bibnamefont {Brevdo}}, \bibinfo {author}
		{\bibfnamefont {Z.}~\bibnamefont {Chen}}, \bibinfo {author} {\bibfnamefont
			{C.}~\bibnamefont {Citro}}, \bibinfo {author} {\bibfnamefont {G.~S.}\
			\bibnamefont {Corrado}}, \bibinfo {author} {\bibfnamefont {A.}~\bibnamefont
			{Davis}}, \bibinfo {author} {\bibfnamefont {J.}~\bibnamefont {Dean}},
		\bibinfo {author} {\bibfnamefont {M.}~\bibnamefont {Devin}}, \bibinfo
		{author} {\bibfnamefont {S.}~\bibnamefont {Ghemawat}}, \bibinfo {author}
		{\bibfnamefont {I.}~\bibnamefont {Goodfellow}}, \bibinfo {author}
		{\bibfnamefont {A.}~\bibnamefont {Harp}}, \bibinfo {author} {\bibfnamefont
			{G.}~\bibnamefont {Irving}}, \bibinfo {author} {\bibfnamefont
			{M.}~\bibnamefont {Isard}}, \bibinfo {author} {\bibfnamefont
			{Y.}~\bibnamefont {Jia}}, \bibinfo {author} {\bibfnamefont {R.}~\bibnamefont
			{Jozefowicz}}, \bibinfo {author} {\bibfnamefont {L.}~\bibnamefont {Kaiser}},
		\bibinfo {author} {\bibfnamefont {M.}~\bibnamefont {Kudlur}}, \bibinfo
		{author} {\bibfnamefont {J.}~\bibnamefont {Levenberg}}, \bibinfo {author}
		{\bibfnamefont {D.}~\bibnamefont {Man\'{e}}}, \bibinfo {author}
		{\bibfnamefont {R.}~\bibnamefont {Monga}}, \bibinfo {author} {\bibfnamefont
			{S.}~\bibnamefont {Moore}}, \bibinfo {author} {\bibfnamefont
			{D.}~\bibnamefont {Murray}}, \bibinfo {author} {\bibfnamefont
			{C.}~\bibnamefont {Olah}}, \bibinfo {author} {\bibfnamefont {M.}~\bibnamefont
			{Schuster}}, \bibinfo {author} {\bibfnamefont {J.}~\bibnamefont {Shlens}},
		\bibinfo {author} {\bibfnamefont {B.}~\bibnamefont {Steiner}}, \bibinfo
		{author} {\bibfnamefont {I.}~\bibnamefont {Sutskever}}, \bibinfo {author}
		{\bibfnamefont {K.}~\bibnamefont {Talwar}}, \bibinfo {author} {\bibfnamefont
			{P.}~\bibnamefont {Tucker}}, \bibinfo {author} {\bibfnamefont
			{V.}~\bibnamefont {Vanhoucke}}, \bibinfo {author} {\bibfnamefont
			{V.}~\bibnamefont {Vasudevan}}, \bibinfo {author} {\bibfnamefont
			{F.}~\bibnamefont {Vi\'{e}gas}}, \bibinfo {author} {\bibfnamefont
			{O.}~\bibnamefont {Vinyals}}, \bibinfo {author} {\bibfnamefont
			{P.}~\bibnamefont {Warden}}, \bibinfo {author} {\bibfnamefont
			{M.}~\bibnamefont {Wattenberg}}, \bibinfo {author} {\bibfnamefont
			{M.}~\bibnamefont {Wicke}}, \bibinfo {author} {\bibfnamefont
			{Y.}~\bibnamefont {Yu}}, \ and\ \bibinfo {author} {\bibfnamefont
			{X.}~\bibnamefont {Zheng}},\ }\href {http://tensorflow.org/} {\enquote
		{\bibinfo {title} {{TensorFlow}: Large-scale machine learning on
				heterogeneous systems},}\ } (\bibinfo {year} {2015}),\ \bibinfo {note}
	{software available from tensorflow.org}\BibitemShut {NoStop}%
	\bibitem [{\citenamefont {Agarap}(2018)}]{agarap2018deep}%
	\BibitemOpen
	\bibfield  {author} {\bibinfo {author} {\bibfnamefont {A.~F.}\ \bibnamefont
			{Agarap}},\ }\href@noop {} {\enquote {\bibinfo {title} {Deep learning using
				rectified linear units (relu)},}\ } (\bibinfo {year} {2018}),\ \Eprint
	{http://arxiv.org/abs/1803.08375} {arXiv:1803.08375 [cs.NE]} \BibitemShut
	{NoStop}%
	\bibitem [{\citenamefont {Kingma}\ and\ \citenamefont
		{Ba}(2014)}]{kingma2014adam}%
	\BibitemOpen
	\bibfield  {author} {\bibinfo {author} {\bibfnamefont {D.~P.}\ \bibnamefont
			{Kingma}}\ and\ \bibinfo {author} {\bibfnamefont {J.}~\bibnamefont {Ba}},\
	}\href@noop {} {\enquote {\bibinfo {title} {Adam: A method for stochastic
				optimization},}\ } (\bibinfo {year} {2014}),\ \Eprint
	{http://arxiv.org/abs/1412.6980} {arXiv:1412.6980 [cs.LG]} \BibitemShut
	{NoStop}%
	\bibitem [{\citenamefont {Lindhard}\ and\ \citenamefont
		{Scharff}(1961)}]{Lindhard1961}%
	\BibitemOpen
	\bibfield  {author} {\bibinfo {author} {\bibfnamefont {J.}~\bibnamefont
			{Lindhard}}\ and\ \bibinfo {author} {\bibfnamefont {M.}~\bibnamefont
			{Scharff}},\ }\href {\doibase 10.1103/PhysRev.124.128} {\bibfield  {journal}
		{\bibinfo  {journal} {Phys. Rev.}\ }\textbf {\bibinfo {volume} {124}},\
		\bibinfo {pages} {128} (\bibinfo {year} {1961})}\BibitemShut {NoStop}%
	\bibitem [{\citenamefont {Lindhard}\ \emph {et~al.}(1963)\citenamefont
		{Lindhard}, \citenamefont {Scharff},\ and\ \citenamefont
		{Schioett}}]{Lindhard1963}%
	\BibitemOpen
	\bibfield  {author} {\bibinfo {author} {\bibfnamefont {J.}~\bibnamefont
			{Lindhard}}, \bibinfo {author} {\bibfnamefont {M.}~\bibnamefont {Scharff}}, \
		and\ \bibinfo {author} {\bibfnamefont {H.~E.}\ \bibnamefont {Schioett}},\
	}\href {https://www.osti.gov/biblio/4153115} {\bibfield  {journal} {\bibinfo
			{journal} {Mat. Fys. Medd. Dan. Vid. Selsk.}\ }\textbf {\bibinfo {volume}
			{33}},\ \bibinfo {pages} {1} (\bibinfo {year} {1963})}\BibitemShut {NoStop}%
	\bibitem [{\citenamefont {Sigmund}(1983)}]{Sigmund1983LSS}%
	\BibitemOpen
	\bibfield  {author} {\bibinfo {author} {\bibfnamefont {P.}~\bibnamefont
			{Sigmund}},\ }\href {\doibase 10.1088/0031-8949/28/3/001} {\bibfield
		{journal} {\bibinfo  {journal} {Physica Scripta}\ }\textbf {\bibinfo {volume}
			{28}},\ \bibinfo {pages} {257} (\bibinfo {year} {1983})}\BibitemShut
	{NoStop}%
	\bibitem [{\citenamefont {Sigmund}(1998)}]{SIGMUND1998}%
	\BibitemOpen
	\bibfield  {author} {\bibinfo {author} {\bibfnamefont {P.}~\bibnamefont
			{Sigmund}},\ }\href {\doibase https://doi.org/10.1016/S0168-583X(97)00638-1}
	{\bibfield  {journal} {\bibinfo  {journal} {Nuclear Instruments and Methods
				in Physics Research Section B: Beam Interactions with Materials and Atoms}\
		}\textbf {\bibinfo {volume} {135}},\ \bibinfo {pages} {1 } (\bibinfo {year}
		{1998})}\BibitemShut {NoStop}%
	\bibitem [{\citenamefont {M.A.}\ and\ \citenamefont {B.Sc.}(1905)}]{Bragg1905}%
	\BibitemOpen
	\bibfield  {author} {\bibinfo {author} {\bibfnamefont {W.~H.~B.}\
			\bibnamefont {M.A.}}\ and\ \bibinfo {author} {\bibfnamefont {R.~K.}\
			\bibnamefont {B.Sc.}},\ }\href {\doibase 10.1080/14786440509463378}
	{\bibfield  {journal} {\bibinfo  {journal} {The London, Edinburgh, and Dublin
				Philosophical Magazine and Journal of Science}\ }\textbf {\bibinfo {volume}
			{10}},\ \bibinfo {pages} {318} (\bibinfo {year} {1905})}\BibitemShut
	{NoStop}%
	\bibitem [{\citenamefont {J.~Lindhard}(1968)}]{Lindhard1968}%
	\BibitemOpen
	\bibfield  {author} {\bibinfo {author} {\bibfnamefont {M.~S.}\ \bibnamefont
			{J.~Lindhard}, \bibfnamefont {V.~Nielsen}},\ }\href@noop {} {\bibfield
		{journal} {\bibinfo  {journal} {Matemat. Fysis. Meddel.}\ }\textbf {\bibinfo
			{volume} {36}},\ \bibinfo {pages} {1} (\bibinfo {year} {1968})}\BibitemShut
	{NoStop}%
	\bibitem [{\citenamefont {Jin}\ \emph {et~al.}(2013)\citenamefont {Jin},
		\citenamefont {Zhang}, \citenamefont {Xue}, \citenamefont {Zhu},\ and\
		\citenamefont {Weber}}]{JIN201365}%
	\BibitemOpen
	\bibfield  {author} {\bibinfo {author} {\bibfnamefont {K.}~\bibnamefont
			{Jin}}, \bibinfo {author} {\bibfnamefont {Y.}~\bibnamefont {Zhang}}, \bibinfo
		{author} {\bibfnamefont {H.}~\bibnamefont {Xue}}, \bibinfo {author}
		{\bibfnamefont {Z.}~\bibnamefont {Zhu}}, \ and\ \bibinfo {author}
		{\bibfnamefont {W.}~\bibnamefont {Weber}},\ }\href {\doibase
		https://doi.org/10.1016/j.nimb.2013.02.051} {\bibfield  {journal} {\bibinfo
			{journal} {Nuclear Instruments and Methods in Physics Research Section B:
				Beam Interactions with Materials and Atoms}\ }\textbf {\bibinfo {volume}
			{307}},\ \bibinfo {pages} {65 } (\bibinfo {year} {2013})}\BibitemShut
	{NoStop}%
	\bibitem [{\citenamefont {Grande}\ \emph {et~al.}(1988)\citenamefont {Grande},
		\citenamefont {Fichtner}, \citenamefont {Behar},\ and\ \citenamefont
		{Zawislak}}]{GRANDE198817}%
	\BibitemOpen
	\bibfield  {author} {\bibinfo {author} {\bibfnamefont {P.}~\bibnamefont
			{Grande}}, \bibinfo {author} {\bibfnamefont {P.}~\bibnamefont {Fichtner}},
		\bibinfo {author} {\bibfnamefont {M.}~\bibnamefont {Behar}}, \ and\ \bibinfo
		{author} {\bibfnamefont {F.}~\bibnamefont {Zawislak}},\ }\href {\doibase
		https://doi.org/10.1016/0168-583X(88)90093-6} {\bibfield  {journal} {\bibinfo
			{journal} {Nuclear Instruments and Methods in Physics Research Section B:
				Beam Interactions with Materials and Atoms}\ }\textbf {\bibinfo {volume}
			{35}},\ \bibinfo {pages} {17 } (\bibinfo {year} {1988})}\BibitemShut
	{NoStop}%
\end{thebibliography}
%

\end{document}